\documentclass[prb,aps,preprint,showpacs]{revtex4}
\usepackage{graphicx}

\begin{document}

\title{First principles lattice dynamics of NaCoO$_2$}

\author{Zhenyu Li}
\author{Jinlong Yang} \thanks{Corresponding author. E-mail: jlyang@ustc.edu.cn}
\author{J.G. Hou}
\author{Qingshi Zhu}

\affiliation{Laboratory of Bond Selective Chemistry and Structure
Research Laboratory, University of Science and Technology of
China, Hefei, Anhui 230026, P.R. China}

\date{\today}

\begin{abstract}

We report first principles linear response calculations on
NaCoO$_2$. Phonon frequencies and eigenvectors are obtained
throughout the Brillouin zone for two geometries with different Na
site occupancies. While most of the phonon modes are found to be
unsensitive to the Na site occupancy, there are two modes
dominated by out-of-plane vibrations of Na giving very different
frequencies for different geometries. One of these two modes, the
A$_{2u}$ mode, is infrared-active, and can be used as a suitable
sensor of Na distribution/ordering. The longitudinal-transverse
splitting of the zone-center optical-mode frequencies, Born
effective charges and the dielectric constants are also reported,
showing considerable anisotropy. The calculated frequencies of
Raman-active modes generally agree with the experimental values of
corresponding Na de-intercalated and/or hydrated compounds, while
it requires better experimental data to clarify the
infrared-active mode frequencies.

\end{abstract}

\pacs{63.20.Dj, 71.20.Ps, 74.25.Kc}

\maketitle

\section{INTRODUCTION}

Since the discovery of the superconductivity in Na$_x$CoO$_2
\cdot$yH$_2$O,\cite{takada03} significant interest has been
focused on this novel material, because there are some intriguing
similarities between this cobalt oxide superconductor and copper
oxide superconductors.\cite{badding03} A number of evidences in
experiments suggest that the superconductivity in this material is
unconventional.\cite{sakurai03,ishida03,higemoto03} Theoretical
models such as resonating valence bond
(RVB)\cite{baskaran03,kumar03,honerkamp03,wang03} and spin-triplet
superconductivity\cite{tanaka03,singh03,ikeda03,tanaka03b} with
strong magnetic quantum fluctuations are also proposed. Although
now it is a consensus that the cobalt oxide superconductor is not
simply a BCS superconductor, some experiments still suggest the
possible strong lattice coupling to the electronic density of
states in this compound.\cite{lupi03,milne04} It is thus important
to fully characterize the lattice dynamical properties of
Na$_x$CoO$_2 \cdot$yH$_2$O and related materials.

While there are some Raman,\cite{iliev04,lemmens03}
infrared,\cite{lupi03,wang03b} and neutron
scattering\cite{boothroyd03} experiments involving the lattice
dynamics of the cobalt oxide compound, there is no first
principles theoretical report found in the literature. In this
article, we report a density functional perturbation theory
(DFPT)\cite{baroni01,gonze97} calculation on NaCoO$_2$, the
intrinsic (without hydration or de-intercalation of Na) insulating
phase of the cobalt oxide superconductor. As shown in Figures 1a
and 1b, NaCoO$_2$ has a hexagonal structure (space group \#194,
P6$_3$/mmc) consisting of CoO$_2$ layers of edge sharing CoO$_6$
octahedra and Na layers with two partly occupied Na Wickoff sites.
To accommodate for the partial occupation of the two Na sites, we
consider two geometries of NaCoO$_2$ with one site fully occupied
and the other site empty, namely geometry $A$ (Figure 1a) with the
$2b$ site occupied and geometry $B$ (Figure 1b) with the $2d$ site
occupied.

\section{COMPUTATIONAL METHOD}

The present results have been obtained through the use of the
ABINIT code,\cite{gonze02} a common project of the Universit¨¦
Catholique de Louvain, Corning Incorporated, and other
contributors (URL http://www.abinit.org), based on
pseudopotentials and plane waves. It relies on an efficient fast
Fourier transform algorithm for the conversion of wave functions
between real and reciprocal spaces, on the adaptation to a fixed
potential of the band-by-band conjugate gradient
method\cite{payne92}, and on a potential-based conjugate-gradient
algorithm for the determination of the self-consistent
potential.\cite{gonze96} Technical details on the computation of
responses to atomic displacements and homogeneous electric fields
can be found in Ref. \onlinecite{gonze97}, and the subsequent
computation of dynamical matrices, Born effective charges,
dielectric permittivity tensors, and interatomic force constants
was described in Ref. \onlinecite{gonze97b}.

Troullier-Martins norm conserving
pseudopotentials\cite{troullier91} are used, with Teter
parametrization\cite{goedecker96} of the Ceperley-Alder
exchange-correlation potential. The kinetic energy cutoff of the
plane wave bases is 50 hartree. A uniform 6$\times$6$\times$2
$k$-point mesh is used in Brillouin-zone integrations, which gives
well converged total energies and phonon frequencies.

\section{RESULTS AND DISCUSSION}

\subsection{Geometries and Electronic Structures}

As a first step, we optimize the geometry and calculate the
corresponding electronic structure. During the geometry
optimization, the cell parameters are fixed to the experimental
values of Na$_{0.74}$CoO$_2$($a=2.84 \AA$,
$c=10.811\AA$).\cite{balsys96} We notice that neutron scattering
experiments\cite{balsys96,jorgensen03,lynn03} for samples with
different Na concentrations give similar cell parameters.
Considering the space group symmetry, the only freedom to be
relaxed is the Co-O layer space $d$, which turns out to be 1.897
and 1.899 \AA\ for geometries $A$ and $B$ respectively. The
electronic band structure and density of states (DOS) based on the
optimized structure of geometry $A$ are shown in Figure 1c, a gap
of 0.94 eV (from 0.60 to 1.54 eV) between the valence band and the
conduction band can be clearly identified. Little difference on
the electronic structure of the two geometries is found. This
similarity strongly indicates that Na contributes to the
electronic structure only by doping electrons to the CoO layers.

\subsection{Phonon Dispersions}

To map the phonon dispersion curves throughout the Brillouin zone,
the dynamical matrices are obtained on a uniform
6$\times$6$\times$2 grid of $q$ points, and real-space force
constants are then found by Fourier transform of the dynamical
matrices. The dynamical matrix at an arbitrary wave vector
\textbf{q} can then be computed by an inverse Fourier transform.
The acoustic sum rule is applied to force the three acoustic
phonon frequencies at $\Gamma$ equal to zero strictly as being
implied by the translation symmetry. The calculated phonon
dispersions and corresponding DOS for geometries $A$ and $B$ are
shown in Figure 2. We see first that both structures are stable.
Second, we notice that the phonon modes are separated to two
groups in frequency, namely the soft group with frequencies below
than about 400 cm$^{-1}$ and the hard group with frequencies
between 450 to 650 cm$^{-1}$. As a whole, the phonon dispersions
for the two geometries with different Na site occupancies are
similar, but there are also some minor differences exist for the
two geometries in the phonon band structure, unlike the nearly
identical electronic band structures. First, as we will discussed
in detail below, there are some soft modes giving significant
different frequencies near zone center for the two geometries.
Secondly, we notice that the dispersion along $\Gamma - A$ for
geometry $B$ is larger than that for geometry $A$.

\subsection{zone-center phonon modes}

The zone-center phonon modes are of special importance, since they
can be obtained by various of experimental methods. In Table I, we
list the frequencies, symmetries and vibration modes of the
zone-center phonons. The triply-degenerated acoustic mode with
zero frequency is not listed. The frequency differences between
the two geometries are generally small, especially for the
Raman-active phonon modes. However there are an infrared-active
A$_{2u}$ mode and a silent B$_{1g}$ mode giving very different
frequencies for geometry $A$ and geometry $B$, and both of these
two modes are mainly contributed by the Na out-of-plane
vibrations. The other A$_{2u}$ and B$_{1g}$ modes (modes 7 and 14)
also involve the out-of-plane vibrations of Na, but we notice that
the vibrations of Co and/or O in these two modes are much stronger
than or at least comparable to the Na vibrations. Therefore they
give relatively small frequency differences for different Na site
occupancies. We find that the frequency difference of Raman-active
A$_{1g}$ mode of O out-of-plane vibrations is rather small, in
contrast to the conclusion of the shell model calculations by
Lemmens et al..\cite{lemmens03} They found a pronounced frequency
dependence of the A$_{1g}$ mode, and argued that it would be used
as a very susceptible sensor of Na distribution/ordering. However,
in our first principles calculations, the infrared-active A$_{2u}$
mode (node 6) in the soft group may be more suitable to act as the
sensor mode. Another difference between $\Gamma$ phonon modes of
the two geometries is the frequency order of the first E$_{2g}$
and B$_{2u}$ modes (modes 1 and 10).

Since the systems we studied here are polar materials, we also get
the longitudinal-transverse optical mode (LO-TO) splittings by
considering response to the macroscopic electric field. As shown
in Table I, the LO-TO splittings are found only in the
infrared-active $2E_{1u}+2A_{2u}$ modes. The biggest LO-TO
splitting is found in the hard  E$_{1u}$ mode, with $\Delta
\omega=\omega_{LO}-\omega_{TO}$ equal to 52.4 and 48.6 cm$^{-1}$
for geometries $A$ and $B$ respectively. Contrastingly, the soft
E$_{1u}$ mode is split by only 5.1 and 5.9 cm $^{-1}$
respectively. Both these two E$_{1u}$ modes are "layer sliding"
modes, and the hard one is mainly contributed by the Co layers
sliding against the O layers. While for the soft E$_{1u}$ mode,
the sliding of the Na layers is about an order stronger than the
relative sliding of Co and O layers. Therefore our results follow
the trend that displacements modulating the "covalent" bonding
produce the largest LO-TO splitting, as being pointed out by Lee
et al.\cite{lee03}.

We also get the Born (dynamical) effective charge tensor. In
hexagonal symmetry, it is diagonal and reduces to two values
$Z_{xx}=Z_{yy}=Z_{//}$ and $Z_{zz}=Z_\perp$. As listed in Table
II, these charges show considerable anisotropy in NaCoO$_2$. We
notice that $Z_\perp$ is much larger than $Z_{//}$ for Na, which
is consistent with the fact that the LO-TO splitting of the soft
in-plane E$_{1u}$ mode is much smaller than that of the soft
out-of-plane A$_{2u}$ mode, with both modes being dominated by the
vibrations of Na. Despite the anisotropy, we define the average
effective charge $\bar{Z}$ as $\frac{1}{3}TrZ$, which gives about
+1, +2 and -$\frac{3}{2}$ for Na, Co, and O, respectively. The
dynamical charges of Co and O are both smaller than their nominal
ionic charges. In addition, the static electronic dielectric
constants are calculated to be $\epsilon^{//}_{\infty}=9.72$,
$\epsilon^{\perp}_{\infty}=4.34$ for geometry $A$ and
$\epsilon^{//}_{\infty}=9.78$, $\epsilon^{\perp}_{\infty}=4.27$
for geometry $B$ respectively.

There are some experimental data on the zone-center phonon
frequencies of Na$_x$CoO$_2$ or Na$_x$CoO$_2 \cdot$yH$_2$O. In
principle, these frequencies can not be directly compared with our
results, since the systems we studied here is NaCoO$_2$. But as we
have seen, the Na site occupancy may not affect the frequencies
much, especially for the high-frequency in-plane vibration modes.
So we also list the experimental Raman and infrared frequencies in
Table I for reference. For NaCoO$_2$ with only one Na site
occupied, symmetry analysis leads to the Raman-active modes
$\Gamma_{Raman}=A_{1g}+E_{1g}+2E_{2g}$ and the infrared-active
optical phonon modes $\Gamma_{IR}=2A_{2u}+2E_{1u}$. The
experimental Raman frequencies listed in Table I generally agree
with our calculated frequencies well. Among the four
infrared-active modes, only the frequency of the hard E$_{1u}$
mode is reported in experiments. The frequency of
Na$_{0.57}$CoO$_2$ reported by Lupi et al.\cite{lupi03} is near
our calculated TO frequency of E$_{1u}$ mode. It is very strange
that infrared spectra by Wang et al.\cite{wang03b} gave four
different infrared frequencies for doubly-degenerated in-plane
E$_{1u}$ mode in metallic Na$_{0.7}$CoO$_2$, where no LO-TO
splitting exists. There should be only two frequencies even
considering the partial Na occupation of the $2b$ and $2d$ site at
the same time. We argue that part of these frequencies may be
contributed by defects or grain surface, since the IR reflectivity
data are very sensitive to surface treatment. In a word, the IR
modes remain to be verified and understood. The peak of optical
phonon at 161.3 cm$^{-1}$ (20 meV) in neutron inelastic scattering
experiment\cite{boothroyd03} is near the frequency of the soft
E$_{2g}$ mode (mode 1) here.

\section{CONCLUSION}

In conclusion, we have calculated the lattice dynamics of
NaCoO$_2$. Most of the phonon modes are only little affected by
the Na site occupancy, and the calculated zone-center phonon
frequencies generally agree with those by recent Raman and neutron
scattering spectra on corresponding Na de-intercalated and/or
hydrated compounds well. Therefore our calculations may also be
helpful for understanding the lattice dynamics of the cobalt oxide
superconductor. However, to clarify the phonon contribution in the
superconductivity requires much further work on the lattice
dynamics and electron-phonon interaction of the superconductor
itself. Experimental data on infrared-active modes are difficult
to compare with our calculated results now. In addition to phonon
frequencies, other useful data such as Born effective charges and
dielectric constants of NaCoO$_2$ are also presented.

\begin{acknowledgments}

This work is partially supported by the National Project for the
Development of Key Fundamental Sciences in China (G1999075305,
G2001CB3095), by the National Natural Science Foundation of China
(50121202, 20025309, 10074058), by the Foundation of Ministry of
Education of China, and by ICTS, CAS.

\end{acknowledgments}

\clearpage

\begin{table}
\caption{Zone center optical phonon modes of NaCoO$_2$. The three
groups separated by horizontal lines are Raman-active,
infrared-active and silent modes respectively. The involved atoms
in each mode and their vibration directions (// for in-plane,
$\perp$ for out-of-plane) are listed. The atomic displacements in
parentheses are obtained by dividing the normalized eigenvector
components by the square root of the atomic mass, and with the
unit of $10^{-2}$ for convenience of reading. These displacements
are only refer to geometry $A$, since they are very similar to
those of geometry $B$. $f^A$ and $f^B$ stand for calculated
frequencies for geometries $A$ and $B$ respectively, and $f^{EXP}$
stands for experimental frequencies. For the four infrared-active
modes, both the LO and TO frequencies are listed. }\label{freq}
\begin{tabular}{cccccccl}
\hline\hline
Mode & Symmetry & \multicolumn{3}{c}{Vibrations}  & $f^A$(cm$^{-1}$) & $f^B$(cm$^{-1}$) & $f^{EXP}$(cm$^{-1}$) \\
\hline
1 & E$_{2g}$ & Na$_{//}$(0.34) &            & O$_{//}$(0.04) & 172.9 & 185.7 & \\
2 & E$_{1g}$ &                 &            & O$_{//}$(0.29) & 477.1 & 482.1 & 458$^a$, 480$^b$, 469$^c$ \\
3 & E$_{2g}$ & Na$_{//}$(0.05) &            & O$_{//}$(0.29) & 483.7 & 489.8 & 494$^a$ \\
4 & A$_{1g}$ &                 &            & O$_\perp$(0.29)& 608.0 & 604.6 & 574$^a$,598$^b$,582$^c$,588$^d$\\
\hline
5 & E$_{1u}$ & Na$_{//}$(0.31) & Co$_{//}$(0.09) & O$_{//}$(0.05) & 201.2 & 216.5 & \\
  &          &                     &             &                & 206.3 & 220.4 & \\
6 & A$_{2u}$ & Na$_\perp$(0.31)& Co$_\perp$(0.09)& O$_\perp$(0.05)& 397.5 & 337.0 & \\
  &          &                     &             &                & 418.1 & 361.6 & \\
7 & A$_{2u}$ & Na$_\perp$(0.03)& Co$_\perp$(0.12)& O$_\perp$(0.24)& 569.8 & 566.5 & \\
  &          &                     &             &                & 618.7 & 615.1 & \\
8 & E$_{1u}$ & Na$_{//}$(0.03) & Co$_{//}$(0.12) & O$_{//}$(0.24) & 586.7 & 590.0 & 570$^e$, (505, 530, 560, 575)$^f$ \\
  &          &                     &             &                & 638.5 & 638.6 &  \\
\hline
9  & E$_{2u}$ &                  & Co$_{//}$(0.18) & O$_{//}$(0.16) &  88.0 &  95.2 & \\
10 & B$_{2u}$ &                  & Co$_\perp$(0.18)& O$_\perp$(0.15)& 197.0 & 172.1 & \\
11 & B$_{1g}$ & Na$_\perp$(0.33) &                 & O$_\perp$(0.10)& 351.7 & 309.9 & \\
12 & E$_{2u}$ &                  & Co$_{//}$(0.12) & O$_{//}$(0.24) & 582.8 & 585.6 & \\
13 & B$_{2u}$ &                  & Co$_\perp$(0.11)& O$_\perp$(0.25)& 616.2 & 610.0 & \\
14 & B$_{1g}$ & Na$_\perp$(0.11) &                 & O$_\perp$(0.28)& 622.4 & 616.0 & \\
\hline\hline
\end{tabular}
\begin{minipage}{6000pt}
\flushleft$^a$ Raman frequency\cite{iliev04} of Na$_{0.7}$CoO$_2$.
\\ $^b$ Raman frequency\cite{lemmens03} of Na$_x$CoO$_2$$\cdot$
$y$H$_2$O powder samples with $x=0.35$. \\
$^c$ Raman frequency\cite{lemmens03} of
Na$_{0.3}$CoO$_2$$\cdot$$1.3$H$_2$O single crystals. \\
$^d$ Raman frequency\cite{lemmens03} of Na$_{0.7}$CoO$_2$ single
crystals. \\
$^e$ Infrared frequency\cite{lupi03} of Na$_{0.57}$CoO$_2$. \\
$^f$ Infrared frequency\cite{wang03b} of Na$_{0.7}$CoO$_2$.
\end{minipage}
\end{table}

\clearpage

\begin{table}
\caption{The Born effective charges of NaCoO$_2$ atoms. See text
for the definition of Z$_{//}$, Z$_\perp$ and $\bar{Z}$.
Superscripts $A$ and $B$ refer to data for geometries $A$ and $B$.
}\label{born}
\begin{tabular}{cccc}
\hline\hline
& Z$_{//}$ & Z$_\perp$ & $\bar{Z}$ \\
\hline
Z$^A$(Na) &  0.87 &  1.37 &  1.03 \\
Z$^A$(Co) &  2.49 &  0.87 &  1.95 \\
Z$^A$(O)  & -1.68 & -1.12 & -1.49 \\
\hline
Z$^B$(Na) &  0.85 &  1.40 &  1.04 \\
Z$^B$(Co) &  2.41 &  0.78 &  1.86 \\
Z$^B$(O)  & -1.63 & -1.09 & -1.45 \\
\hline\hline
\end{tabular}
\end{table}

\clearpage
\begin{figure}
\caption{(color online) (a) Structure model of geometry $A$. The
smallest balls represent cobalt atoms, and the balls in red
bonding to Co are oxygen atoms. Between CoO layers, there exist
sodium (big purple ball) layers. (b) Structure model of geometry
$B$. (c) Electronic band structure and density of states of
NaCoO$_2$.} \label{banddos}
\includegraphics{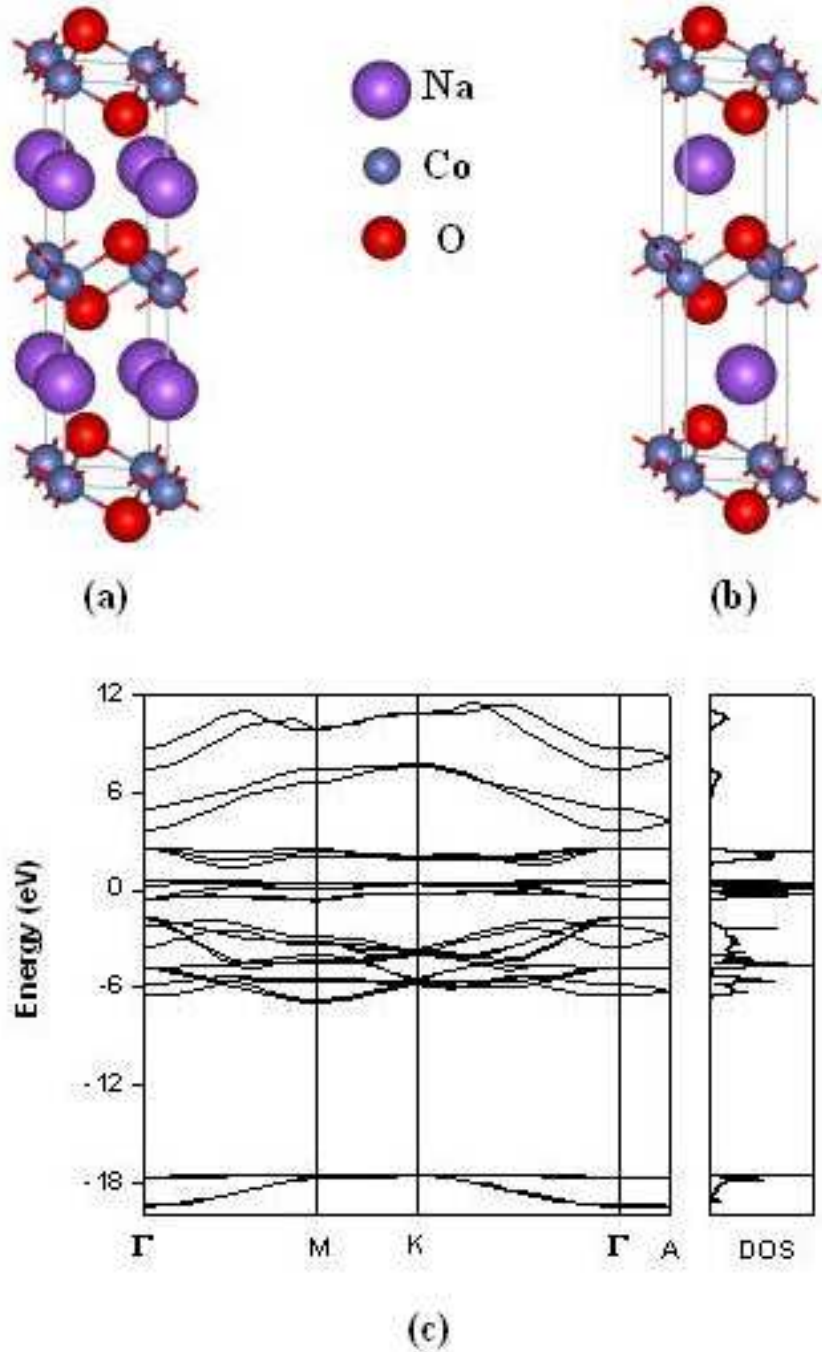}
\end{figure}

\clearpage
\begin{figure}
\caption{Phonon band structure and density of states of NaCoO$_2$
for (a) geometry $A$ and (b) geometry $B$ respectively.}
\label{banddosph}
\includegraphics{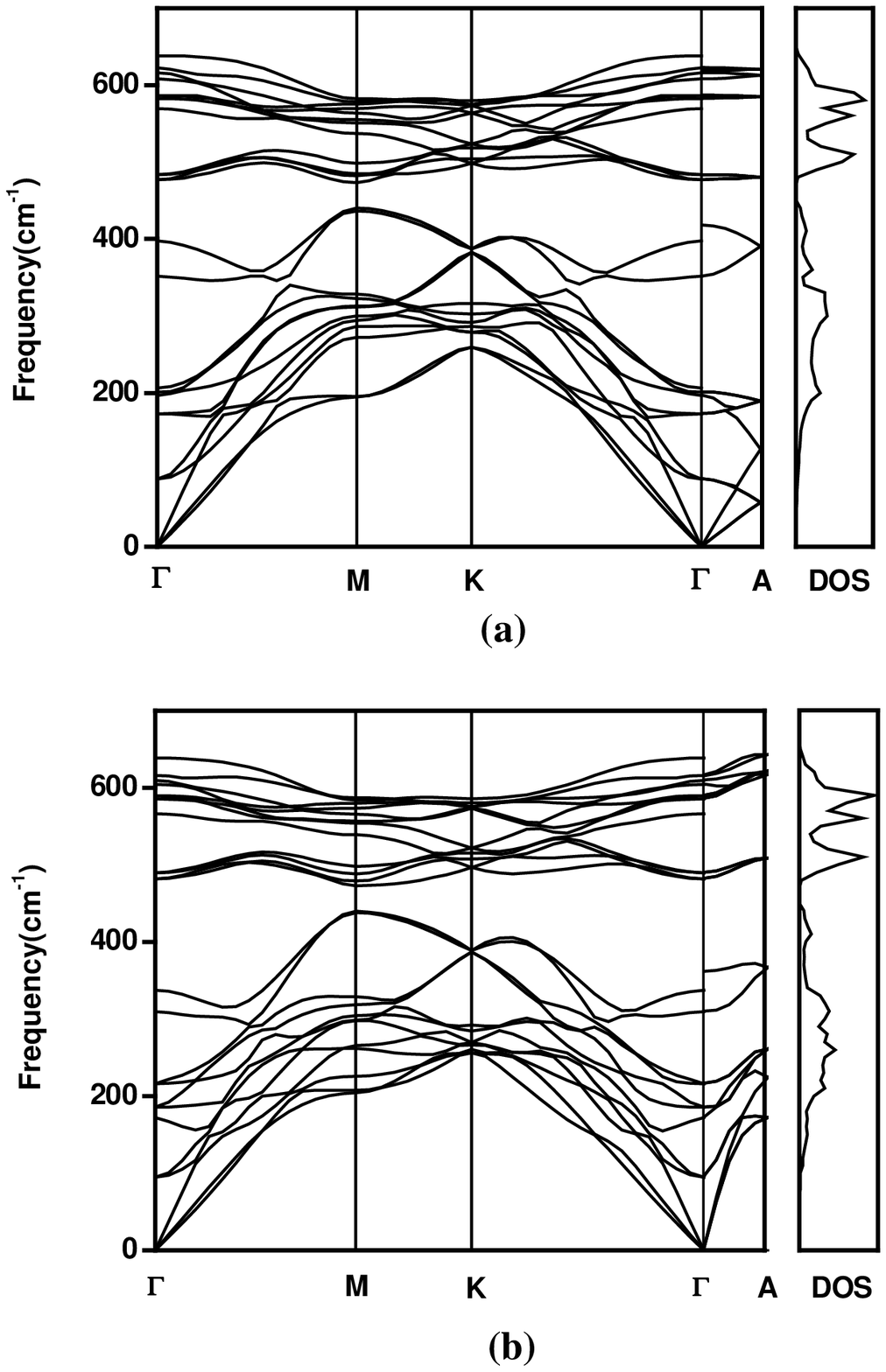}
\end{figure}

\end{document}